\documentclass{ws-procs975x65}

\usepackage{graphicx}

\begin{document}

\title{STUDYING THE STOCHASTICITY IN ASTROPHYSICAL SIGNALS}

\author{EMIL POGHOSIAN}

\address{ Yerevan Physics Institute and Yerevan State University, Yerevan, Armenia\\
E-mail: emil@yerphi.am}

\begin{abstract}
The efficiency of the Kolmogorov-Arnold technique for the astrophysical signals is studied modeling sequences with both random and regular properties. This technique has been applied to the study of the structures in cosmic microwave background maps obtained by the Wilkinson Microwave Anisotropy Probe.
\end{abstract}


\bodymatter

\section*{}
The study of the Cosmic Microwave Background (CMB) temperature maps has been performed by various methods
\cite{Hinshaw}. Kolmogorov stochasticity parameter (K-parameter)  \cite{Kolm,Arnold} has been among those methods \cite{KSP1,KSP2} which enabled the study of non-Gaussian structures such as the Cold Spot  in the data obtained by WMAP; the behavior of the Kolmogorov's parameter for the Cold Spot is compatible to the void nature of that anomaly.

Kolmogorov stochasticity parameter \cite{Kolm,Arnold} $\lambda_n$ defined for a sequence of real numbers
\begin{equation}
\lambda_n = \sqrt{n}\, sup|F(X)-F_n(X)|,
\label{eq:lambda}
\end{equation}
where $F(X)$ and $F_n(X)$ are the theoretical and empirical distribution functions, respectively, is determined by a distribution which at $n \rightarrow \infty $ tends uniformly to a universal function
\begin{equation}
\Phi(\lambda) = \sum_{k=-\infty}^{+\infty}{(-1)^k e^{-2k^2\lambda^2}},\\
\,\,\Phi(0) = 0,
\label{eq:phi}
\end{equation}
independent on $F$. 
This parameter has been applied by Arnold to find out the degree of randomness of given progressions and systems following from the number theory (cf.\cite{GP}). 

We studied the behavior of Kolmogorov's parameter generating 400 different sequences of given length divided into four groups, each of 100 sequences each \cite{Tigran}: I. Gaussian distribution; II. the same with perturbation; III. Gaussian distribution extracted from a sequences $x_n=a$ $n$ mod $b$, where $a$ and $b$ are constants; IV. two copies of the same Gaussian sequence.

\begin{figure}[!htbp]
	\centering
		\includegraphics[width=2.3 in]{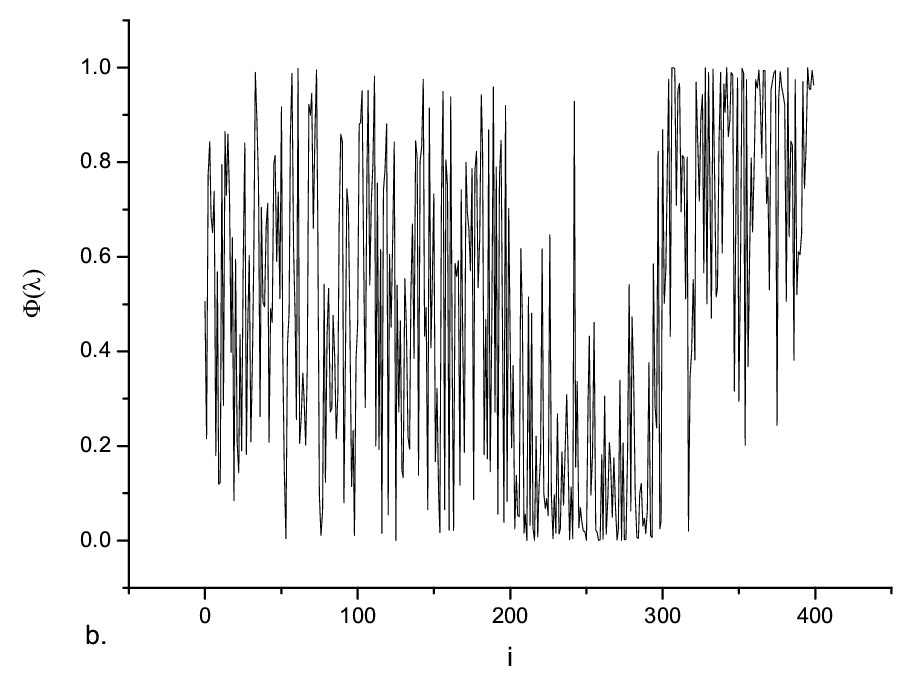}
		\includegraphics[width=2.3 in]{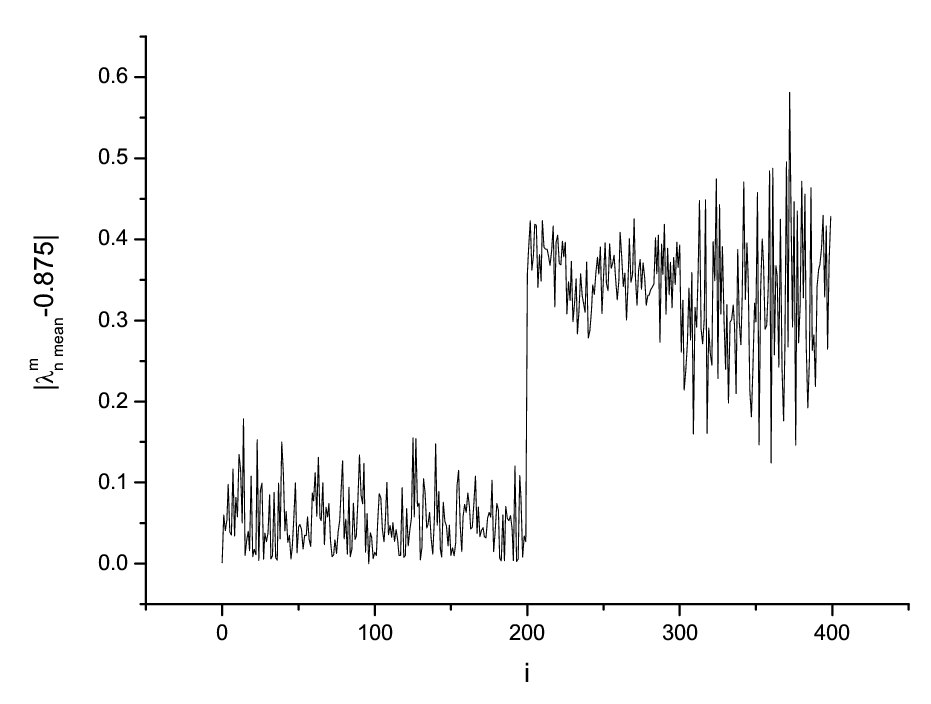}
		\caption{Computed values of $\Phi(\lambda)$  and $|\lambda^n_{mean}-\lambda_m|$.}
	\label{fig:test_1}
\end{figure}
 
For these groups of sequences the values of Kolmogorov's function $\Phi(\lambda)$ and the absolute value of differences of the empirical and theoretical values of $\lambda_{mean}$ have been computed, where the mean theoretical value of Kolmogorov stochasticity parameter yields
\begin{equation}
\lambda_{m}=\int{\lambda\phi(\lambda)d\lambda}\approx 0.869, \,\, \phi(\lambda)=\Phi'(\lambda).
\end{equation}

These computations, as shown in Fig. 1, enabled to find out the behavior of $\lambda$ and the distribution $\Phi$ for the modeled sequences i.e. with initially known properties and then consider the CMB as a composition of signals of stochastic and regular properties \cite{KSP2}. Such properties appear to be enough informative to distinguish the contribution of the non-cosmological signals into the CMB, for example, of the radiation of the Galactic disk. Particularly, the Galactic radiation was shown to possess high value of $\lambda$ and $\Phi$ around 1, which indicates its non-Gaussian nature, while CMB signal is basically Gaussian with inhomogeneous perturbations of various degree of randomness. Also, this approach can be helpful in the revealing of the power spectrum of the distribution of inhomogeneities of matter in the Universe, the voids, complementing the data of the large scale surveys \cite{Perc}.

\end{document}